\documentclass{article}
\usepackage{fixltx2e}
\usepackage{authblk}
\usepackage[latin1]{inputenc}
\usepackage{amsmath,amsfonts,amssymb,bm}
\usepackage{graphicx}
\usepackage{cite}
\usepackage[tight,hang,small,bf]{subfigure}

\usepackage{graphicx,color}% Include figure files

\begin{document}

\title{Fast beam steering with full polarization control using a galvanometric optical scanner and polarization controller}

\author{M.~Jofre,$^{1,*}$~G.~Anzolin,$^{1}$~F.~Steinlechner,$^{1}$~N.~Oliverio,$^{2}$~J.~P.~Torres,$^{1,3}$\\V.~Pruneri,$^{1,4}$~and M.~W.~Mitchell$^{1,4}$}

\affil{\small{$^{1}$ ICFO-Institut de Ciencies Fotoniques, Castelldefels, E-08860 Barcelona, Spain.\\
$^{2}$ Signadyne, Castelldefels, E-08860 Barcelona, Spain.\\
$^{3}$ Dept. Signal Theory and Communications, Universitat Polit\`{e}cnica de Catalunya, E-08034 Barcelona, Spain.\\
$^{4}$ ICREA-Instituci\'{o} Catalana de Recerca i Estudis Avan\c{c}ats, E-08010 Barcelona, Spain.}}

\affil{*marc.jofre@icfo.es} %% email address is required

\date{}

\maketitle

\begin{abstract}Optical beam steering is a key element in many industrial and scientific applications like in material processing, information technologies, medical imaging and laser display. Even though galvanometer-based scanners offer flexibility, speed and accuracy at a relatively low cost, they still lack the necessary control over the polarization required for certain applications. We report on the development of a polarization steerable system assembled with a fiber polarization controller and a galvanometric scanner, both controlled by a digital signal processor board. The system implements control of the polarization decoupled from the pointing direction through a feed-forward control scheme. This enables to direct optical beams to a desired direction without affecting its initial polarization state. When considering the full working field of view, we are able to compensate polarization angle errors larger than $0.2$ rad, in a temporal window of less than $\sim 20$ ms. Given the unification of components to fully control any polarization state while steering an optical beam, the proposed system is potentially integrable and robust.\end{abstract}
%%%%%%%%%%%%%%%%%%%%%%%%%%  body  %%%%%%%%%%%%%%%%%%%%%%%%%%

\section{Introduction}
The steering of free-space optical beams is of great interest in many industrial and scientific applications including material processing, information technologies, medical imaging and laser display \cite{Marshall2004}. Precise and rapid scanning can be accomplished with galvanometric scanners, which deflect a beam from a pair of galvanometer-mounted mirrors. The steering process introduces a polarization rotation which depends in a complex but predictable way on the steering angles. Here we address the issue of polarization control in these scanners and demonstrate a steering system capable of delivering an arbitrary polarization to an arbitrary direction at high speed. The system can also operate in a polarization-transparent mode, in which an arbitrary, unknown input polarization is preserved as the scanner directs the beam. Polarization-controlled steering systems could reduce complexity in existing applications, including ellipsometry \cite{Azzam1977}, polarization optical coherence tomography \cite{Drexler2008}, and distant interferometry with polarized light \cite{Hariharan2003}. It may also enhance the applicability of current systems in quantum communications \cite{Rarity2003,Gisin2007} and polarization-based stereoscopic projectors \cite{Bogaert2008}.

Among the different optical scanning techniques developed so far, galvanometer-based scanners (galvos) offer flexibility, speed and accuracy at a relatively low cost. Current galvo technology achieves closed-loop bandwidths of several kilohertz even for beams with large radii. Moreover, a resolution at the microradian level can be achieved within a large scanning field, which is usually of the order of $\pm 20$\textdegree \cite{Cho2006}. A galvo system is capable of deflecting $s$- and $p$-polarization components of the beam towards the same direction, although, as it generates the scan the corresponding change of the Fresnel coefficients \cite{Born1999} in the two mirrors generate a pointing-dependent polarization state transformation of the output beam.

We report here on the development of a polarization steerable system assembled with a fiber polarization controller (PC) and a galvanometric scanner, both controlled by a digital signal processor (DSP) board. The system implements the control of the polarization state, which is decoupled from the pointing direction of the galvo, by compensating with the PC the polarization transformation induced by the galvo through a feed-forward control scheme. Hence, the method presented can be extended to similar systems where it is needed to decouple the system general state of polarization from other transformations.

Among other applications, our solution is of practical relevance for free-space quantum communication. To date, every free-space quantum key distribution (QKD) system developed has made use of the polarization to encode quantum information. A step forward for exceeding the current transmission length limit, around $200$ km, would be to use satellites as network nodes \cite{Rarity2002,Aspelmeyer2003,Bonato2009}. Satellite-tracking telescopes typically incorporate rapid, fine-pointing subsystems employing rotating mirrors as in our galvanometer system. The dynamic polarization control strategy described here may be also applied for polarization control in these systems \cite{Bonato2006,Bonato2007}.

\section{System overview}
\begin{figure}[htbp]
\centering
\includegraphics[angle=0,width=1\textwidth]{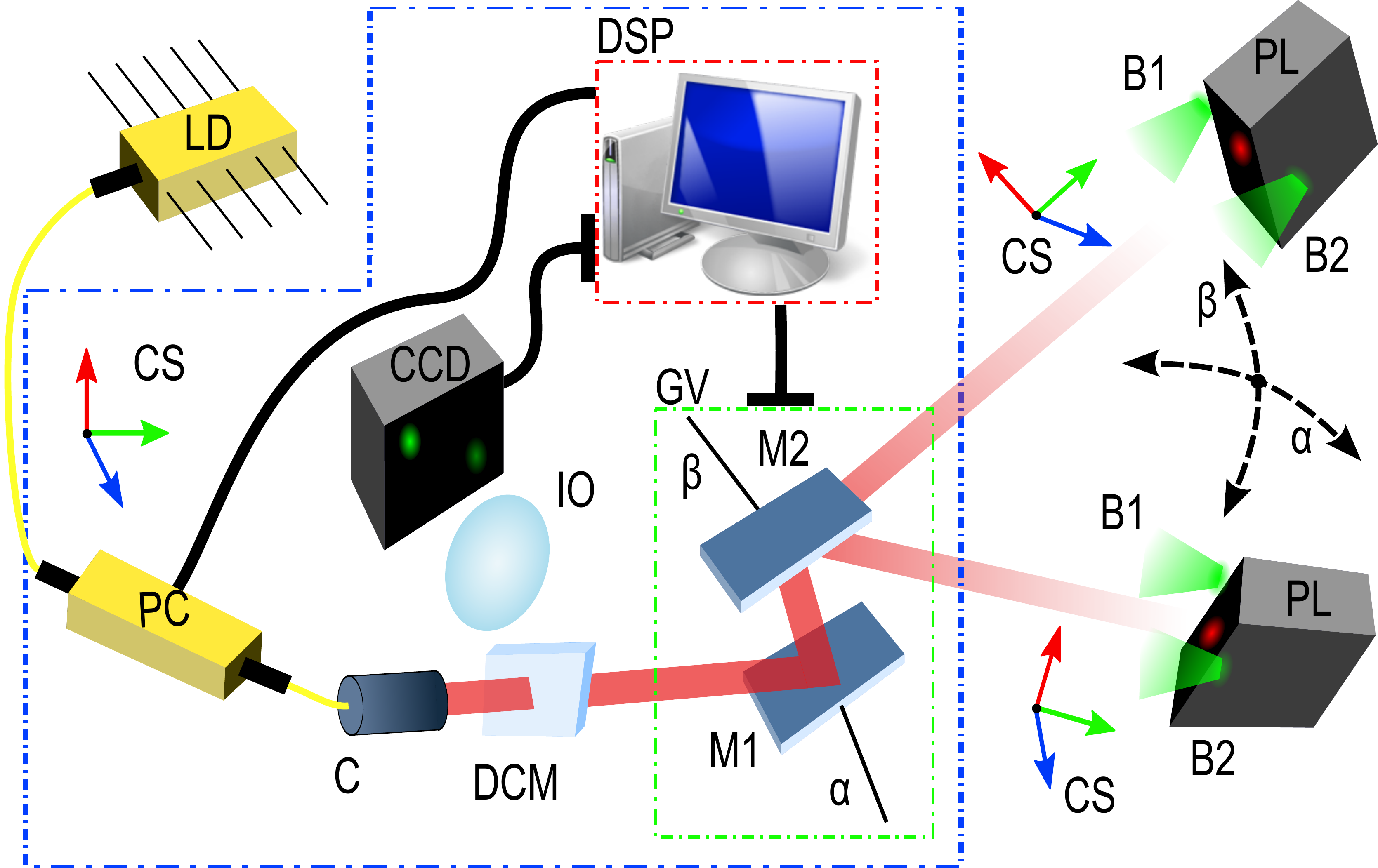}
\caption{Optical system scheme. (LD) denotes the laser diode, (PC) polarization controller, (C) fiber-collimator, (CS) coordinate system, (DCM) dichroic mirror, (IO) imaging optics, (CCD) charge-coupled device camera, (DSP) digital signal processor, (GV) galvanometric scanner with (M1) mirror 1, ($\alpha$) angle M1, (M2) mirror 2, ($\beta$) angle M2, (PL) polarimeter, (B1) beacon 1 and (B2) beacon 2.}
\label{Fig:PC_GV_System_Scheme}
\end{figure}

The integrated system implemented makes use of commercially available components, shown in Fig. \ref{Fig:PC_GV_System_Scheme}. A fiber-pigtailed laser diode (LD) at $850$ nm generates a single polarization state with a high degree of polarization (DOP) into the PC. The output of the PC is properly launched into the galvo system (GV) using a fiber to free-space collimator (C). A dichroic mirror (DCM), placed before the galvo, reflects the signal from the beacons to be imagined on the CCD through an imaging optics system. Finally, the two imaged beacons on the CCD are processed by a digital signal processor (DSP) to compensate relative angle rotations of the receiver with respect to the transmitter, and galvo polarization transformations at a rate of $\sim 20$ ms.

The PC is constituted by a cascade of three piezoelectric actuators which squeeze a single-mode fiber (SMF) along different directions. The expansion or contraction of an actuator in response to the applied driving voltage produces a variable pressure on the SMF and thus a controlled stress-induced birefringence. Fiber PCs based on fiber squeezing with piezoelectric actuators exhibit a number of remarkable performance: low insertion loss ($0.5$ dB), low polarization dependent loss ($<0.1$ dB), and response time of less than $100\mu$s. The PC is driven with a $12$-bit digital signal and the associated driver generates voltages from $0$ V to $140$ V for each of the three actuators. The galvo system consists of two mirrors that provide $\pm 20$\textdegree rotation angles ($\pm40$\textdegree optical), $5$ ms response times, $2.2$ arcsec angular resolution and has $20$ mm mirror diameter. The CCD camera has a frame rate of $108$ fps at full resolution with $10\mu$s exposure time and $8\mu$m$\times8\mu$m square pixels. In front of the CCD camera an imaging optics system with $70$ mm effective focal length is placed to get $3.44$\textdegree field of view (FOV). The aperture of the galvo is large enough to resolve the two beacons with a few cm spatial separation at $2$ m distance.

\section{System operation}
The polarization transformation introduced by the galvo and by the relative Tx-Rx angle of rotation are conveniently described by Stokes-Mueller formalism. The optical properties of a two-axis galvo optical scanner constituted by a pair of rotating planar mirrors are described in \cite{Anzolin2010}. The polarization state matrix description of the system is
\begin{equation}\label{Eq:SystemMatrixDescription}
\vec{S}_{out}=\mathbf{R}\left(\chi\right)\cdot\mathbf{G}\left(\alpha,\beta\right)\cdot\mathbf{PC}\left(\phi_1,\phi_2,\phi_3\right)\cdot\vec{S}_{in}.
\end{equation}
$\mathbf{R}\left(\chi\right)$ is the Tx-Rx orientation matrix, $\mathbf{G}$ is the galvo matrix as a function of the respective the mirror rotation angles $\alpha$ and $\beta$, $\mathbf{PC}$ the equivalent matrix of the PC as a function of the introduced phases of each of the three actuators used to compensate the system polarization change, $\vec{S}_{in}$ and $\vec{S}_{out}$ the input and output Stokes vectors.

\subsection{Galvanometer scanner model}\label{Sec:GalvoScannerModel}
\begin{figure}[htbp]
\begin{center}
\subfigure[Galvo system model.]{\label{Fig:GV_System_Scheme}\includegraphics[angle=0,width=0.5\textwidth]{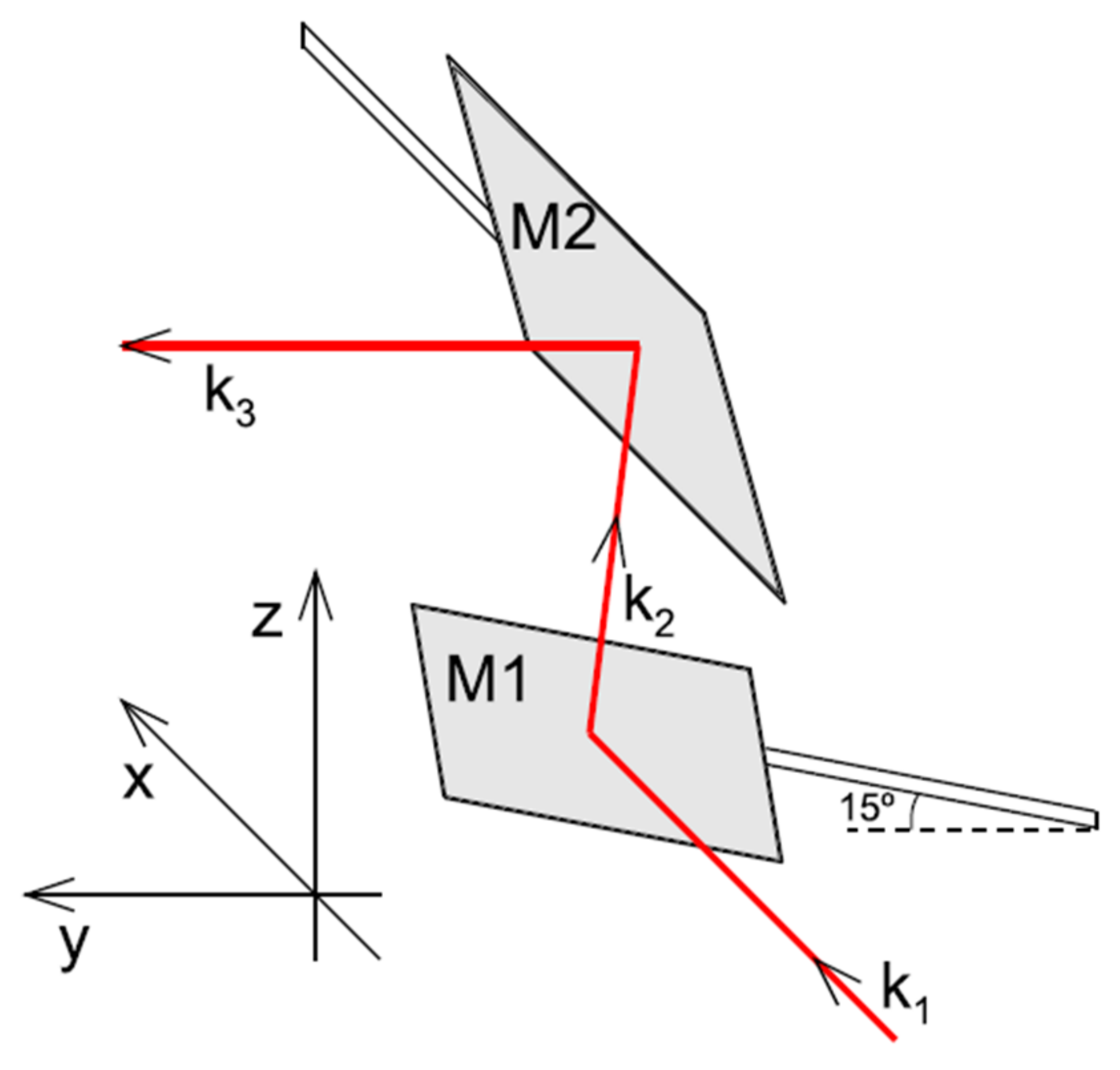}}
\subfigure[Galvo mirror model.]{\label{Fig:GV_Mirror_substrates_Scheme}\includegraphics[angle=0,width=0.45\textwidth]{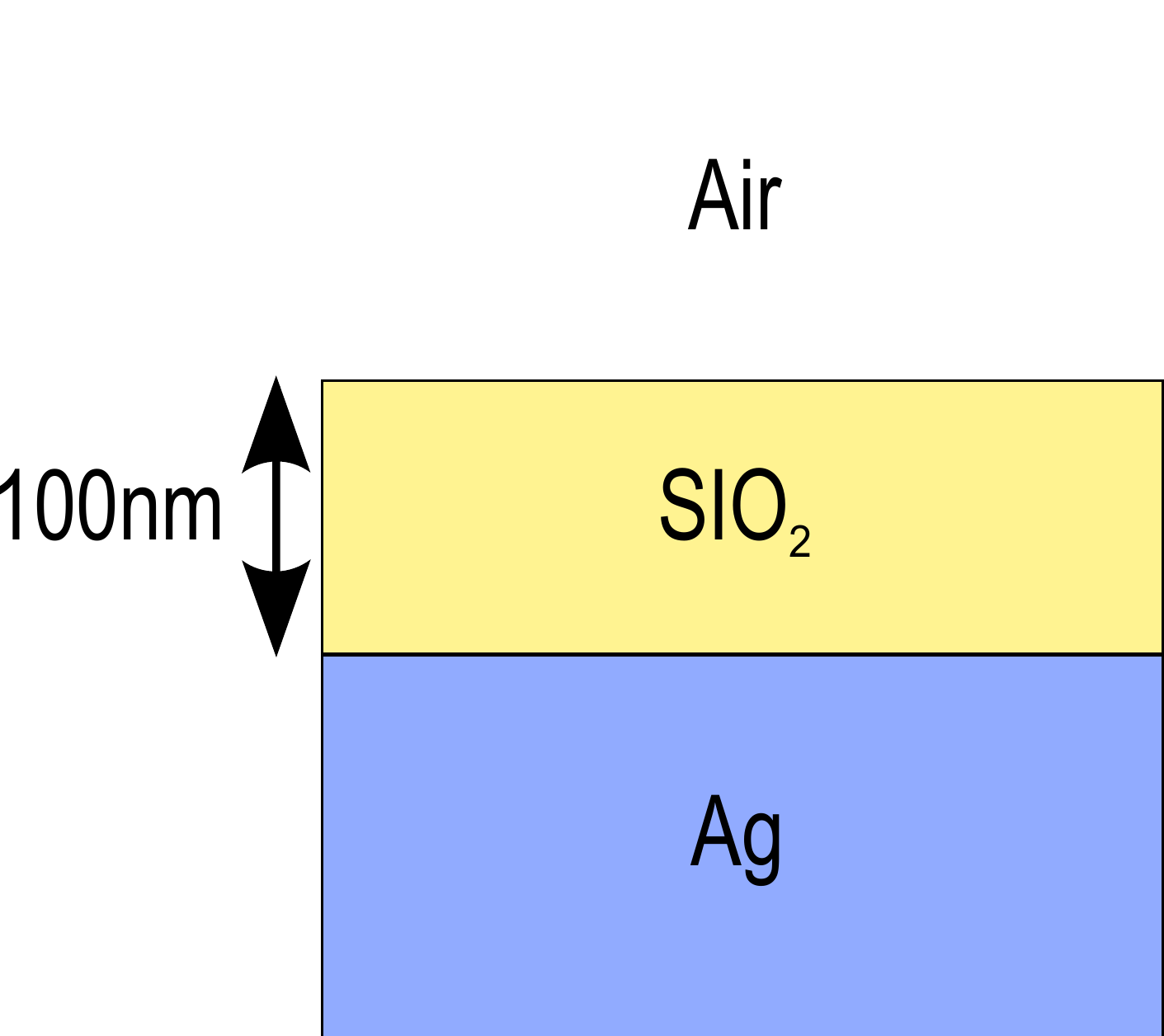}}
\caption{Galvo scanner system model. (a) Scheme of the galvo scanner system, with the reference frame used in the calculations. An example of the optical path of a ray is indicated by the red line. (b) Physical model of the galvo mirrors substrate. The structure of the mirrors considers a quasi-real model of the galvo mirrors two-layer-compound made of a protective dielectric layer of quartz (SIO$_2$), and the reflective substrate of silver (Ag).}
\label{Fig:GV_Scheme}
\end{center}
\end{figure}

The galvo scanner used is assembled with two rotating mirrors as shown in Fig. \ref{Fig:GV_System_Scheme}. The first mirror (M1) is rotated by $45$\textdegree around the $y$-axis and then by an angle $\Gamma_1=15$\textdegree around the $x$-axis. The second mirror (M2), is rotated by an angle $\Gamma_2=45$\textdegree around the $x$-axis. The mirrors of the current system were specifically designed to minimize the difference in loss between $s$- and $p$- polarization components for an operational wavelength of $850$nm. Figure \ref{Fig:GV_Mirror_substrates_Scheme} shows a two-layer model of the galvo mirrors. The layers are made a protective dielectric layer of quartz (SIO$_2$) with $1.43$ refractive index and thickness of $100$nm, and the silver (Ag) reflective substrate with a refractive index of $0.152+5.721i$. The silver layer can be considered to have an infinite thickness due to the short penetration depth of the beam at the operation wavelength.

Considering the impinging angles of the optical beam and the mirrors structure model, the Mueller matrix description of the galvo system consists in the concatenation of a reference-frame rotation and reflection Mueller matrices. First, a reference-frame rotation before impinging on M1, a reflection Mueller matrix on M1 considering the correspondent Fresnel coefficients depending on the angle $\alpha$, a reference-frame rotation before impinging on M2, a reflection matrix on M2 computing the Fresnel coefficient with respect the angle $\beta$ and final reference frame matrix rotation \cite{Anzolin2010}. Using this model, the introduced overall phases and attenuations at the output between $s$- and $p$-polarization components can be computed. The difference in attenuation for the $s$- and $p$-polarization components introduces an effective polarization dependent loss smaller than $0.05$ and is not compensated.

\subsection{Receiver orientation angle}
\begin{figure}[htbp]
\centering
\includegraphics[angle=0,width=1\textwidth]{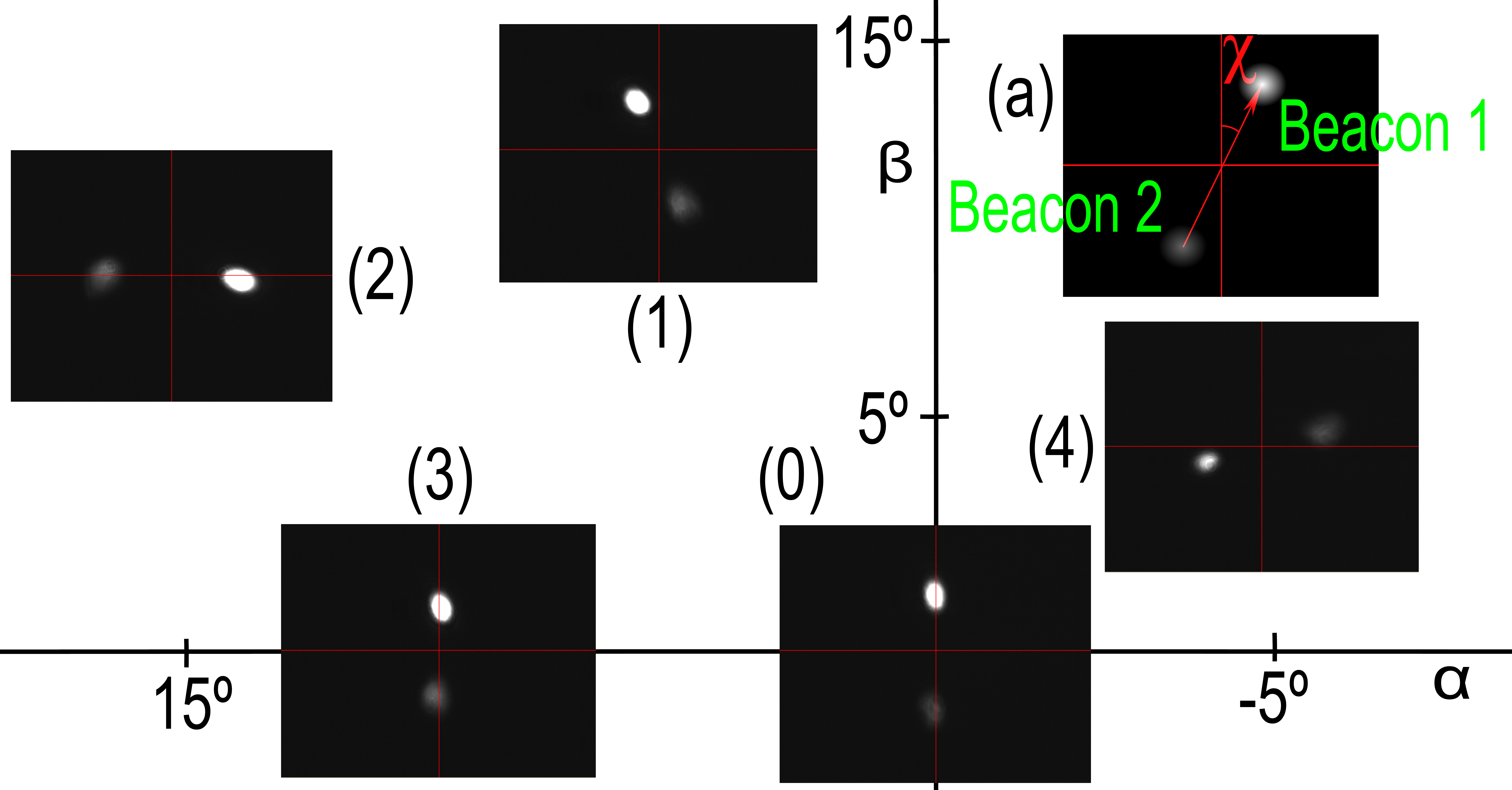}
\caption{Imaged beacons to retrieve the Tx-Rx relative angle rotation $\chi$. (a) The angle is computed between the central vertical axis of the CCD image and the clockwise angle to the brighter beam spot. Notice that one of the beams is brighter than the other to easily identify both spots. Five pointing directions are considered with galvo mirrors' angles and particular receiver orientation rotation, which are grouped in the triplet ($\alpha$,$\beta$,$\chi$): (0) zero pointing corresponds to ($0$\textdegree,$0$\textdegree,$0$\textdegree), (1) pointing 1 ($3.56$\textdegree,$12.19$\textdegree,$336.47$\textdegree), (2) pointing 2 ($18.66$\textdegree,$8.37$\textdegree,$93.06$\textdegree), (3) pointing 3 ($7.99$\textdegree,$-0.13$\textdegree,$4.01$\textdegree) and (4) pointing 4 ($-4.03$\textdegree,$3.30$\textdegree,$243$\textdegree).}
\label{Fig:CCDTwoBeaconsImageCompilation}
\end{figure}

The relative Tx-Rx orientation angle $\chi$ is retrieved by imaging the receiver through the galvanometer system and computing the apparent rotation angle.  This imaging task is facilitated by two beacons, one brighter than the other, mounted on either side of the receiver. Figure \ref{Fig:CCDTwoBeaconsImageCompilation}(a) sketches the angle measurement algorithm. $\chi$ is the angle between the center vertical axis of the CCD image and the clockwise angle to the brighter spot. The rotation $\chi$ is converted into Mueller matrix formalism as Eq. (\ref{Eq:MuellerAngleRotationMatrix}). Notice that the galvo system rotates by $90$\textdegree an image, independently of the pointing direction. In particular, for our demonstration, we consider as reference the zero pointing direction which already involves a $90$\textdegree image rotation. Therefore, for any angle $\chi$ corresponding to the rotation of polarization to apply, relative to the zero pointing direction, we have
\begin{equation}\label{Eq:MuellerAngleRotationMatrix}
\mathbf{R}\left(\chi\right)=
\begin{bmatrix}
1 & 0 & 0 & 0\\
0 & \cos\left(\chi\right) & \sin\left(\chi\right) & 0\\
0 & -\sin\left(\chi\right) & \cos\left(\chi\right) & 0\\
0 & 0 & 0 & 1\\
\end{bmatrix}.
\end{equation}

\subsection{Polarization controller}
\newcommand{\bA}{{\rm \bf A}}
\newcommand{\uphi}{\underline{\phi}}
The transfer function of piezoelectric actuators generally show hysteretic behavior \cite{Hirabayashi2003}, initial loading curve and are rate dependent \cite{Ge1995,Ang2007}. While a highly accurate modeling of these effects would allow precise compensation in arbitrary trajectories \cite{Tan2009}, here we adopt a simple compensation scheme which consists in forcing each actuator to stay in a particular hysteresis cycle previously well calibrated. The cycle is defined by the extreme voltages $V_{min}=0$ V and $V_{max}=75$ V. The first step consists in warming-up the three actuators by injecting sinusoidal voltages peaking at $V_{min}$ and $V_{max}$, to drive the system to the initial state. Subsequent changes always proceed by a return to $0$V, during a short time ($100\mu$s), and then to the final voltage $V$. To overtake the rate dependent behavior, we force the system to change every $20$ ms.

Controlled stress along a SMF induce a variable linear birefringence between parallel and perpendicular components respect to the applied force \cite{Johnson1979,La2011}. Each PC actuator $k$ $\left(k=1,2,3\right)$ is equivalent to a wave plate with a variable phase retard $\phi_k$ and a fixed direction of the fast axis. Hence, the Mueller matrix associated to an actuator is equivalent to the matrix of an elliptical retarder expressed in terms of the axis $\hat{n}_{k}$ and the phase retard between the components along the fast and slow axis $\phi_{k}$ \cite{Li2008}. Furthermore, a preload is present in each actuator so that they are always in contact with the SMF, in order to have a response for any voltage $V>0$. As a result, a non-null phase retard $\uphi_{k}$ is introduced by each actuator for zero voltage. A general description of the PC is given by the concatenation of the three actuators ${\bA}$ as
\begin{equation}\label{eq:PCGeneraldef}
{\rm \bf PC}(\phi_1,\phi_2,\phi_3) = {\bA}(\hat{n}_{3},\phi_3+\uphi_{3}) \cdot {\bA}(\hat{n}_{2},\phi_2+\uphi_{2}) \cdot {\bA}(\hat{n}_{1},\phi_1+\uphi_{1}).
\end{equation}

${\bA}$ corresponds to a polarization rotation around the axis $\hat{n}=\left(n_x,n_y,n_z\right)$ by an angle $\phi$, as
\begin{equation}
{\bA}(\hat{n},\phi) \equiv {\bf I} \cos \phi + {\bf n}_\times \sin \phi + \hat{n} \otimes \hat{n} (1-\cos \phi)
\end{equation}
where $\bf I$ is the identity matrix, 
\begin{equation}
 {\bf n}_\times \equiv 
\begin{bmatrix}
0 & 0 & 0 & 0\\
0 & 0 & - n_z & n_y\\
0 &n_z & 0 & -n_x\\
0 & -n_y& n_x & 0\\
\end{bmatrix}.
\end{equation}
and
\begin{equation}
 \hat{n} \otimes \hat{n} \equiv 
\begin{bmatrix}
1 & 0 & 0 & 0\\
0 & n_x^2 & n_x n_y & n_x n_z\\
0 & n_x n_y & n_y^2 & n_y n_z\\
0 & n_x n_z& n_y n_z & n_z^2\\
\end{bmatrix}.
\end{equation}

Initially $\hat{n}_k$, $\phi_k\left(V\right)$ and $\uphi_k$ are unknown. We note that the system can be described in a more practical formulation where $\uphi_k$ are removed from the rotation matrix ${\bA}(\hat{n},\phi+\uphi) = {\bA}(\hat{n},\phi) {\bA}(\hat{n},\uphi)$, so that 
\begin{equation}
{\rm \bf PC}(\phi_1,\phi_2,\phi_3) = {\bA}(\hat{n}_{3},\phi_3) \cdot {\bA}(\hat{n}_{3},\uphi_{3}) \cdot {\bA}(\hat{n}_{2},\phi_2+\uphi_{2}) \cdot {\bA}(\hat{n}_{1},\phi_1+\uphi_{1}).
\end{equation}
By scanning the voltage $V_3$, we learn $\hat{n}_3$ and $\phi_3(V)$, but not $\uphi_3$. For the remainder of the procedure, we leave $V_3=0$, so that $\phi_3 = 0$, and the rotation ${\bA}(\hat{n}_{3},\phi_3)$ is the identity, giving 
\begin{equation}
{\rm \bf PC}(\phi_1,\phi_2,\phi_3) = \cdot {\bA}(\hat{n}_{3},\uphi_{3}) \cdot {\bA}(\hat{n}_{2},\phi_2+ \uphi_2)  \cdot {\bA}^{-1}(\hat{n}_{3},\uphi_{3}) \cdot {\bA}(\hat{n}_{3},\uphi_{3}) \cdot {\bA}(\hat{n}_{1},\phi_1+\uphi_{1}).
\end{equation}
We note that ${\bA}(\hat{n}_{3},\uphi_{3}) \cdot {\bA}(\hat{n}_{2},\phi_2+ \uphi_2)  \cdot {\bA}^{-1}(\hat{n}_{3},\uphi_{3}) $ is simply another rotation ${\bA}(\hat{n}'_{2},\phi_2+ \uphi_2)$, about another unknown axis $\hat{n}'_2$. As before, we sweep $V_2$ to find $\hat{n}'_2$ and $\phi_2(V)$ but not $\uphi_2$. We set $\phi_2=0$ for the next step, and find 
\begin{equation}
{\rm \bf PC}(\phi_1,\phi_2,\phi_3) =  {\bA}(\hat{n}''_{1},\phi_1+\uphi_{1}) \cdot {\bA}(\hat{n}'_{2},\uphi_{2})  \cdot {\bA}(\hat{n}_{3},\uphi_{3})
\end{equation}
where $\hat{n}''_1$ is the new axis for the first actuator, considering the rotations by both $\uphi_3$ and $\uphi_2$. After finding $\hat{n}''_1$ and $\phi_1(V)$,
\begin{equation}
{\rm \bf PC}(\phi_1,\phi_2,\phi_3) = {\bA}(\hat{n}_{3},\phi_3) \cdot {\bA}(\hat{n}'_{2},\phi_2) \cdot {\bA}(\hat{n}''_{1},\phi_1) \cdot {\bA}(\hat{n}''_{1},\uphi_{1})\cdot {\bA}(\hat{n}'_{2},\uphi_{2})  \cdot {\bA}(\hat{n}_{3},\uphi_{3}).
\end{equation}
The last three terms, which are unknown, can be combined into a single rotation Mueller matrix $\mathbf{M}$, which is independent of drive voltages. Thus, Eq. (\ref{eq:PCGeneraldef}) is equivalently described as,
\begin{equation}
{\rm \bf PC}(\phi_1,\phi_2,\phi_3) = {\bA}(\hat{n}_{3},\phi_3) \cdot {\bA}(\hat{n}'_{2},\phi_2) \cdot {\bA}(\hat{n}''_{1},\phi_1) \cdot \mathbf{M}.
\end{equation}

Note that Eq. (\ref{eq:PCGeneraldef}) describes an arbitrary polarization rotation, and thus suffices to describe the effect of the polarization controller including possible rotations due to birefringence in the SM input and output fibers.

\subsubsection{Differential phase calibration}
\begin{figure}[htbp]
\centering
\includegraphics[angle=0,width=1\textwidth]{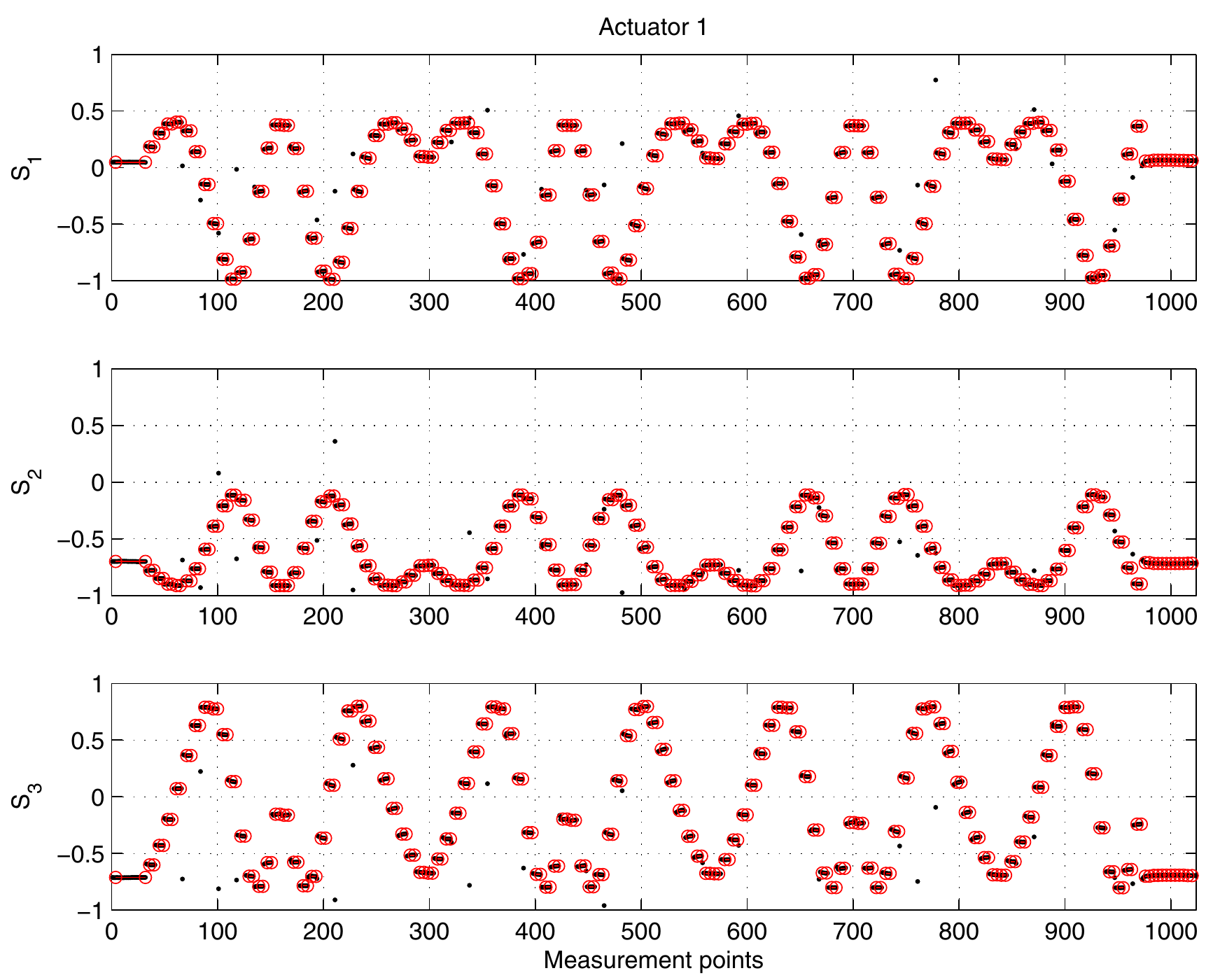}
\caption{Stokes parameters measured while voltage-scanning actuator 1. The recorded Stokes parameters sequence follows a squared cosine function. It is repeated three and a half times to acquire several cycles through a specific PC actuator. The data is periodically sampled asynchronously. Each red circle corresponds to an identified voltage and Stokes parameters measurement pair. Black points that do not fit the pattern correspond to data samples at the return instant to $0$V of the PC driving strategy. (top) $S_1$ , (middle) $S_2$ and (bottom) $S_3$ Stokes parameters. Similar measurement records are obtained for actuators 2 and 3 of the PC.}
\label{Fig:StokesMeasurementSerieGalvoActuator1}
\end{figure}

The calibration process for the PC consists in recording the output Stokes parameters with the polarimeter at zero pointing while a given PC actuator is voltage scanned from $0$ V to $75$ V and back for several cycles while the remaining actuators remain at $0$ V. Zero pointing corresponds to the pointing direction of the galvo for zero angle rotation of its mirrors. Figure \ref{Fig:StokesMeasurementSerieGalvoActuator1} shows the measured Stokes parameters through the voltage scan of PC actuator 1. Similar measurement records are obtained for the other two actuators of the PC. 

\begin{figure}[htbp]
\centering
\includegraphics[angle=0,width=1\textwidth]{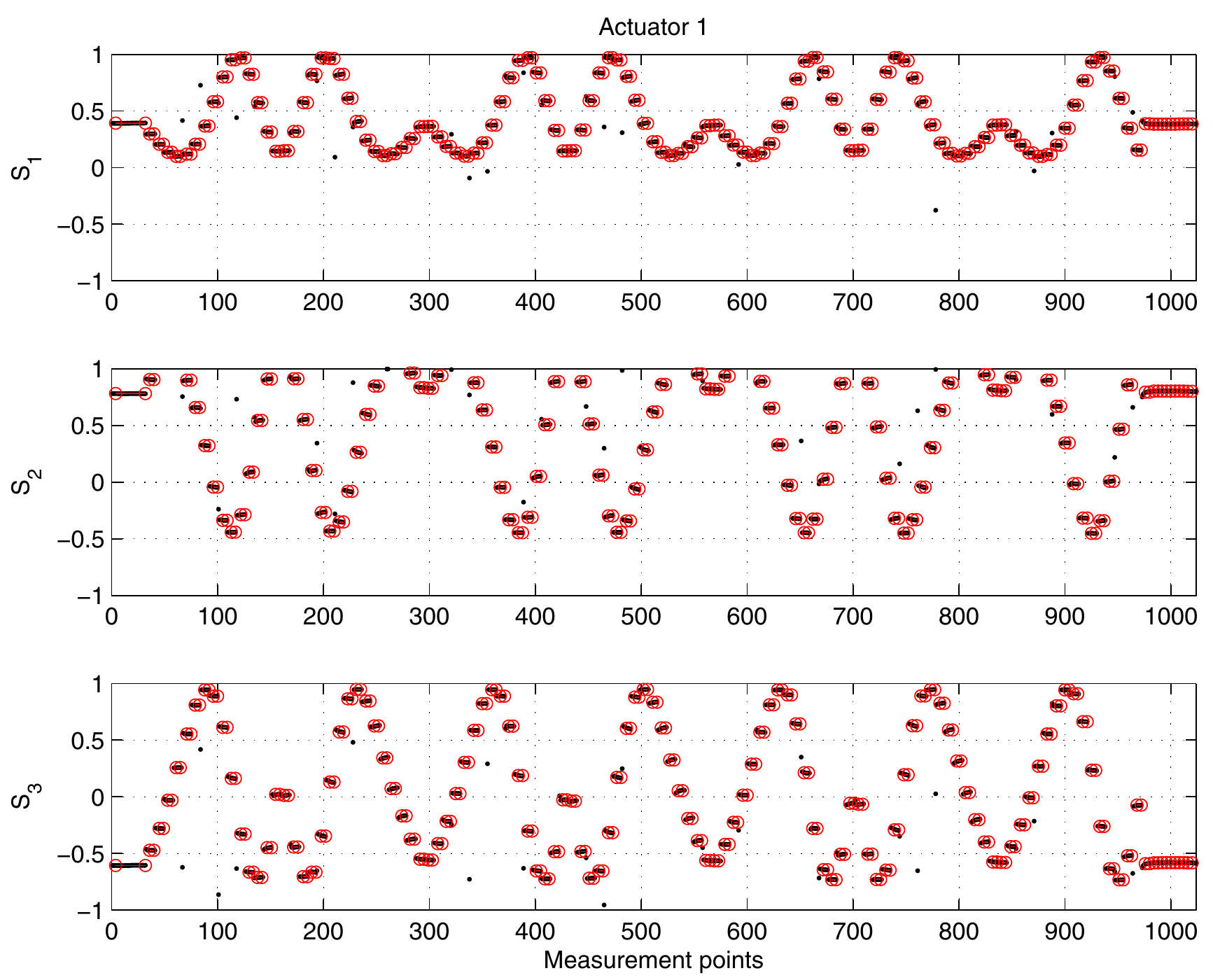}
\caption{Processed Stokes parameters obtained by removing from the raw data the polarization transformation contribution of the galvo. To remove the galvo contribution, the inverse of the galvo Mueller matrix for zero pointing is applied. (top) $S_1$ , (middle) $S_2$ and (bottom) $S_3$ Stokes parameters. Similar measurement records are obtained for actuators 2 and 3 of the PC.}
\label{Fig:StokesMeasurementSerieActuator1}
\end{figure}

\begin{figure}[htbp]
\centering
\includegraphics[angle=0,width=1\textwidth]{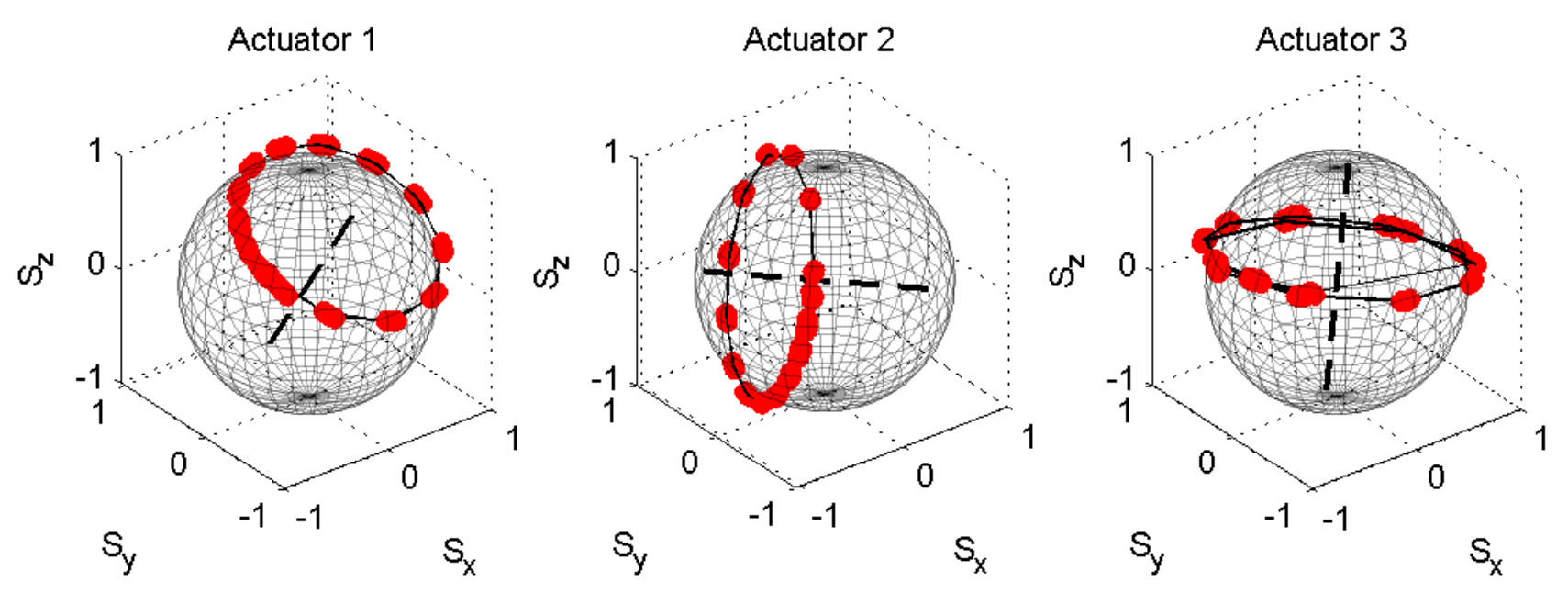}
\caption{Polarization state measurement, from the PC data, on the Poincare sphere and associated actuator axis $\hat{n}_k$ for each of the three PC actuators. For the current measurement the different $\left(\hat{n}_3,\hat{n}'_2,\hat{n}''_1\right)$ are oriented in azimuth and elevation ($\theta$,$\varphi$) as follows: ($29.52$\textdegree,$9.61$\textdegree) actuator 1, ($122.88$\textdegree,$-6.22$\textdegree) actuator 2 and ($45.15$\textdegree,$43.35$\textdegree) actuator 3. The first and the third actuators are similarly oriented along the same direction, while the second actuator is at $90$\textdegree with respect to them.}
\label{Fig:MeasAroundPrincipalAxisActuators}
\end{figure}

\begin{figure}[htbp]
\centering
\includegraphics[angle=0,width=1\textwidth]{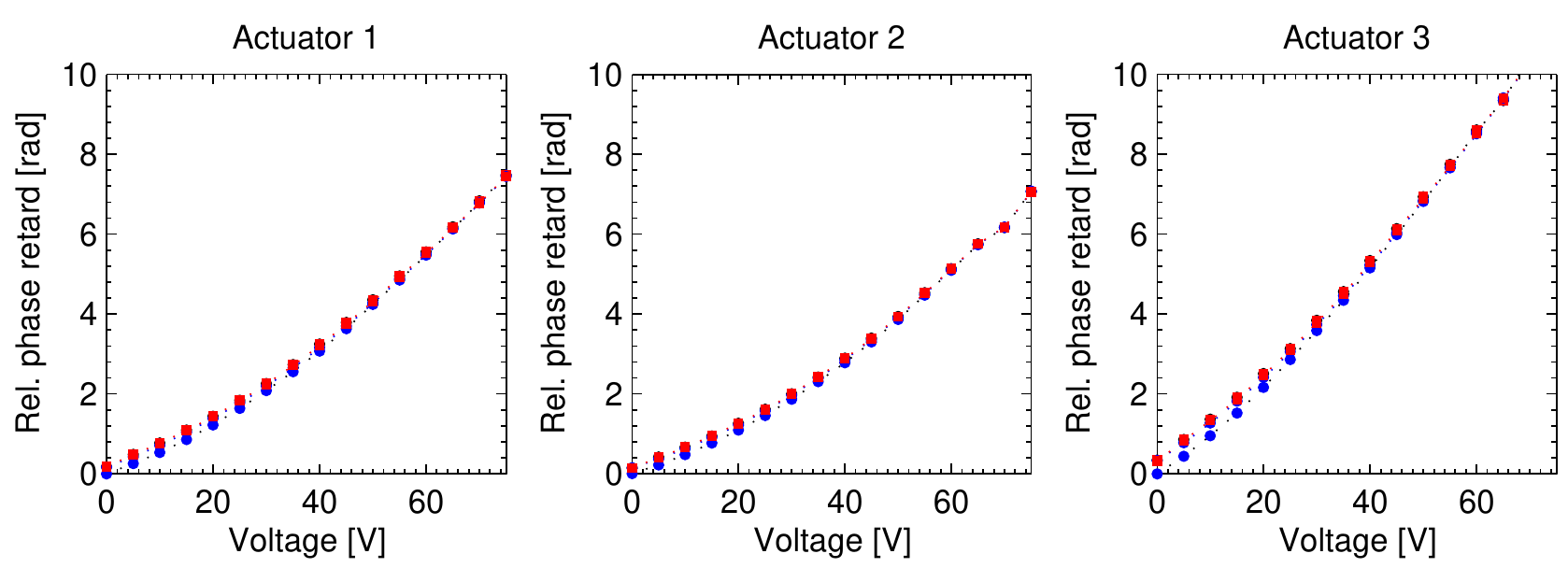}
\caption{Introduced relative phase retard with respect to the applied driving voltage for each PC actuator. Hysteresis has been strongly reduced by returning the drive voltage to zero before each measurement, and relatively slow driving speeds. The phase retard retrieved is relative to $0$ V driving voltage. Blue circles (red squares) show the phase trajectory with increasing (decreasing) voltage.}
\label{Fig:VoltageToPhaseActuator}
\end{figure}

\begin{figure}[htbp]
\centering
\includegraphics[angle=0,width=1\textwidth]{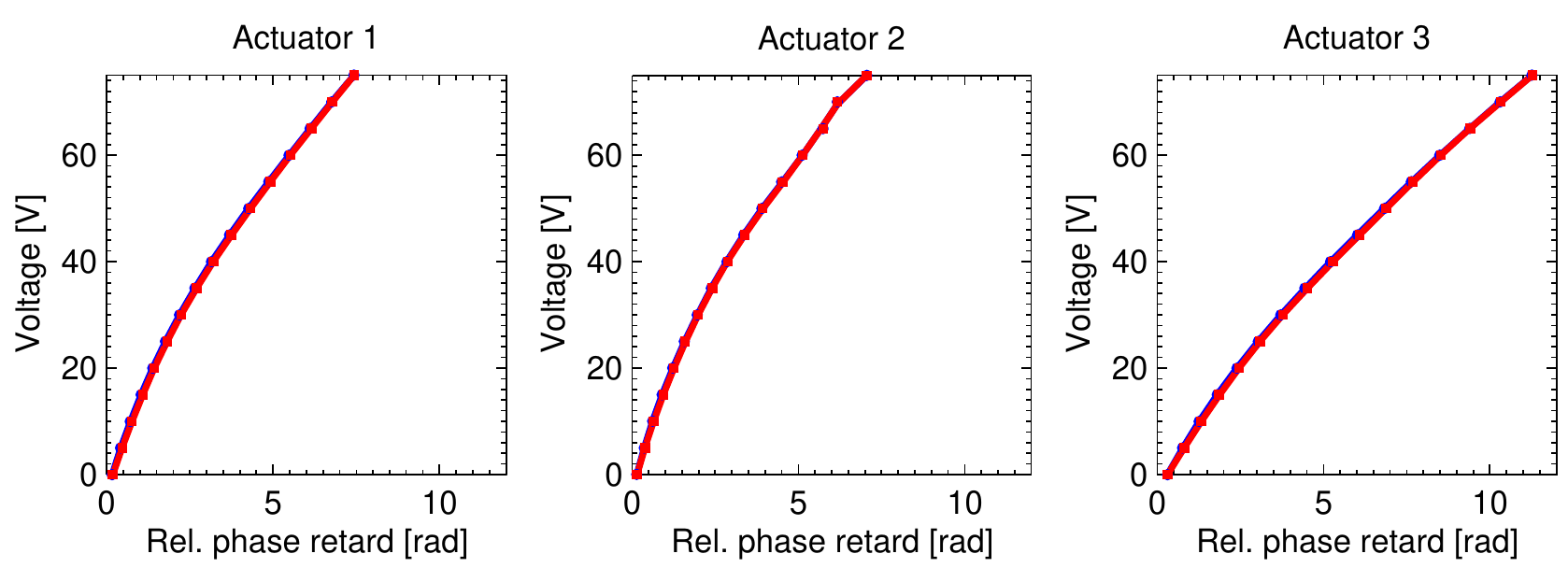}
\caption{Polarization controller inferred transfer function. Phase to voltage relation for each of the three PC actuators. The ascending and descending curves are each fit with 6th-order polynomials for real-time computation. With proper combination of the three PC phase retards, it is possible to generate any rotation matrix.}
\label{Fig:PhaseToVoltageEstimationActuator}
\end{figure}

The output polarization state includes polarization rotations from the galvo due to the fact that the PL measures the polarization at the output of the system at zero pointing. To remove the galvo contribution, we apply the inverse of the Mueller matrix describing the galvo transformation for zero pointing. For clarification, the measurements with the PL could have been taken for any other pointing direction and consequently the contribution from the galvo to be removed would correspond to the inverse of the galvo polarization transformation for the particular pointing direction used. The galvo Mueller matrix has been constructed analytically considering the two-mirror galvo model described in detail in \cite{Anzolin2010}. The resulting polarizations, corresponding to only the PC transformation, are shown in Fig. \ref{Fig:StokesMeasurementSerieActuator1}. The current data directly allows to retrieve $\left(\hat{n}_3,\hat{n}'_2,\hat{n}''_1\right)$ and the relative phases versus drive voltage $\phi\left(V_k\right)$, computed by the angle subtended between consecutive polarization states, as shown in Fig. \ref{Fig:MeasAroundPrincipalAxisActuators}. This procedure finds the phase relation relative to the applied voltage including hysteresis both when increasing and when decreasing the driving voltage, shown in Fig. \ref{Fig:VoltageToPhaseActuator}. The measured voltage-to-phase relation is inverted computationally to obtain the phase-to-voltage relation, shown in Fig. \ref{Fig:PhaseToVoltageEstimationActuator}. The ascending and descending curves are each fit with 6th-order polynomials.

The calibration procedure compensates also for birefringent and geometrical effects in the input SMF to the PC. The implemented scheme overcomes this rotation by considering extra relative phase retards and PC actuators' $\left(\hat{n}_3,\hat{n}'_2,\hat{n}''_1\right)$ axis rotations while calibrating. In practice, we find these effects to be stable within $0.1$ rad over $10$ minutes. Thus, the calibration parameters are valid within this time window. The basic operation cycle of the system is as follows: first, the PC is calibrated taking approximately $2$ minutes, most probably having the specific application using this system paused; second, once the new calibration parameters are updated, the system can be used for $8$ minutes. Clearly, this basic operation cycle can be repeated continuously for long operation of the system. The operation of the system could be improved by placing a polarimeter control system together with the CCD imaging system, to allow continuous calibration of the PC.

\subsubsection{Calibration of $\mathbf{M}$}
$\mathbf{M}$ is fully determined by a polarization rotation and the phase between non-orthogonal polarization states. This enables to operate the system in a polarization-transparent mode, i.e., any input polarization state is maintained through out the system. The retrieval of $\mathbf{M}$ for polarization-transparent calibration requires the measurement of the Stokes parameters, at the output of the system, for two different but well-known non-orthogonal input polarizations. Such a calibration requires an extra optical device that provides a controlled polarization state at the input, such as a deterministic polarization controller or, if the provided light's polarization is completely fixed, a polarization modulator suffices. Then, knowing the input polarization, and the effect of the polarization controller as a function of ($\left((V_1),(V_2),(V_3)\right)$), we can set ($\left((V_1),(V_2),(V_3)\right)$) to produce the identity for the concatenation of the parts of the system known, $\mathbf{I}=\mathbf{R}\left(\chi\right)\cdot\mathbf{G}\left(\alpha,\beta\right)\cdot{\bA}(\hat{n}_{3},\phi_3(V_3)) \cdot {\bA}(\hat{n}'_{2},\phi_2(V_2)) \cdot {\bA}(\hat{n}''_{1},\phi_1(V_1))$. At present, $\mathbf{M}$ is computed by solving the non-linear system of equations Eq. (\ref{Eq:Msystem}),
\begin{equation}\label{Eq:Msystem}
 \left\{\begin{aligned}
        \vec{S}_{out_1}=\mathbf{I}\cdot\mathbf{M}\left(\theta,\varphi,\delta\right)\cdot\vec{S}_{in_1}\\
     \vec{S}_{out_2}=\mathbf{I}\cdot\mathbf{M}\left(\theta,\varphi,\delta\right)\cdot\vec{S}_{in_2}
       \end{aligned}
       \right.
 \qquad \text{.}
\end{equation}
$\vec{S}_{out_1}$ and $\vec{S}_{out_2}$ are the output polarizations corresponding to the input polarizations $\vec{S}_{in_1}$ and $\vec{S}_{in_2}$.

In our demonstration, we aim at maintaining a given output polarization, measured at zero pointing angle, at any other pointing direction. Therefore, polarization-transparent operation is not strictly needed. This requires only to identify the polarization rotation measuring the transformation of a single input polarization state, which reduces the system Eq. (\ref{Eq:Msystem}) to,
\begin{equation}
\vec{S}_{out_1}=\mathbf{I}\cdot\mathbf{M}\left(\theta,\varphi,\delta\right)\cdot\vec{S}_{in_1}.
\end{equation}

\subsubsection{Polarization compensation}
Any polarization state transformation through the system consists in a rotation of the Poincare sphere. With the proper combination of the three phase retards $\left(\phi_1,\phi_2,\phi_3\right)$ of the PC it is possible to generate any rotation matrix. Hence, the overall polarization rotation compensation which the PC has to introduce can be derived from Eq. (\ref{Eq:SystemMatrixDescription}), and consists in generating the inverse of the galvo and the relative Tx-Rx orientation angles as
\begin{equation}\label{Eq:PCCompensation}
\mathbf{PC}\left(\phi_1,\phi_2,\phi_3\right)=\mathbf{G}\left(\alpha,\beta\right)^{-1}\cdot\mathbf{R}^{-1}\left(\chi\right).
\end{equation}

The three particular phases to achieve in Eq. (\ref{Eq:PCCompensation}) are computed using a non-linear fitting routine which minimizes the square of the difference between the required target matrix, given the PC measured parameters.

\section{Polarization compensation performance}
The mobile polarization analyzer system consists of a polarimeter (PL) with two $639$ nm fiber-collimated laser beacons mounted on opposite sides pointing to the transmitter. The two beacons are imaged using a charge-coupled device camera (CCD) to retrieve any relative angle rotation between the receiver PL and the transmitter galvo. The PL consists of a rotating quarter-wave plate, a fixed polarizer and a photodiode with $\pm 0.25$\textdegree azimuth and ellipticity angle accuracy. The laser beacons are mounted at $60$ mm radius with respect the center aperture of the PL and are collimated to a $1$ mm beam waist.

Five pointing directions were considered to quantify the performance of the system, with arbitrary placements and rotations of the receiver PL. The target pointing directions are described by the angles of the galvo mirrors and receiver orientation rotation, grouped in the triplet ($\alpha$,$\beta$,$\chi$). Figure \ref{Fig:CCDTwoBeaconsImageCompilation}(0) shows the zero pointing which corresponds to ($0$\textdegree,$0$\textdegree,$0$\textdegree), considering it as the reference. The other four pointing directions shown in Fig. \ref{Fig:CCDTwoBeaconsImageCompilation} are: (1) pointing 1 ($3.56$\textdegree,$12.19$\textdegree,$336.47$\textdegree), (2) pointing 2 ($18.66$\textdegree,$8.37$\textdegree,$93.06$\textdegree), (3) pointing 3 ($7.99$\textdegree,$-0.13$\textdegree,$4.01$\textdegree) and (4) pointing 4 ($-4.03$\textdegree,$3.30$\textdegree,$243$\textdegree).

A manual polarization controller was added before the PC in order to generate different input polarization states, taking them as reference at the zero pointing. Four extra different pointing directions with arbitrary receiver rotations have been considered, taking four different polarization states measurements at each pointing direction. The error angle $\Delta \epsilon$ is defined as the absolute arc angle between the final polarization state and the reference polarization state at the zero pointing direction as
\begin{equation}\label{Eq:FormulaAngleError}
\Delta\epsilon=\left|\frac{1}{2}\cos^{-1}\Big(\cos\left[2\left(\theta_{px}-\theta_{pz}\right)\right]\cos\left[2\left(\varphi_{px}-\varphi_{pz}\right)\right]\Big)\right|.
\end{equation}
$\theta_{px}$ and $\varphi_{px}$ are azimuth and ellipticity for a particular final pointing direction, and $\theta_{pz}$ and $\varphi_{pz}$ are azimuth and ellipticity for the zero pointing, considering an initial polarization state.

\begin{figure}[htbp]
\centering
\includegraphics[angle=0,width=0.55\textwidth]{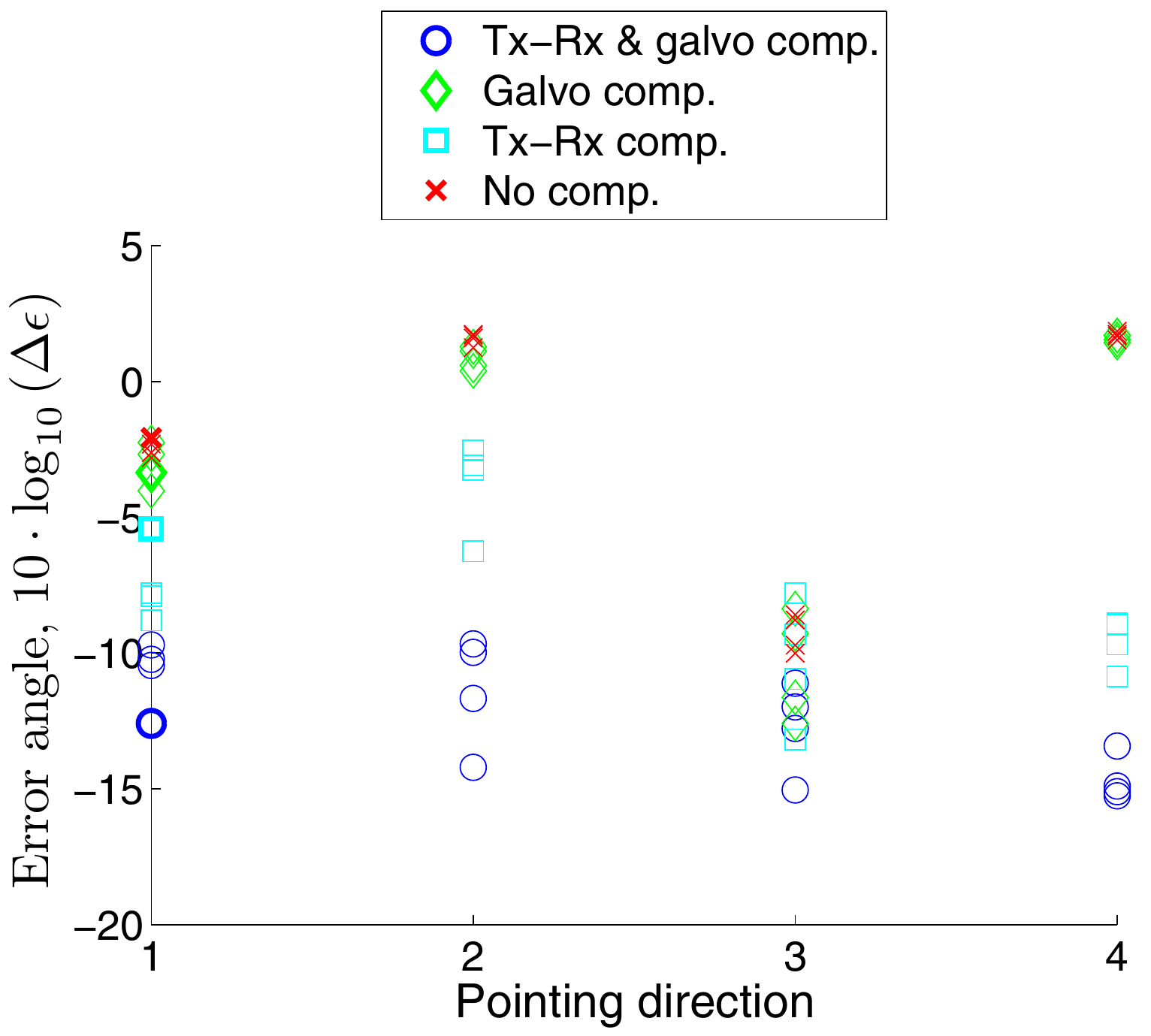}
\caption{Error angle for different polarization states. The polarization states are measured with the receiving polarimeter. Different compensation configurations are presented, compensation of the Tx-Rx and galvo (blue-circle marker), galvo compensation (green-diamond marker), Tx-Rx orientation (cyan-square marker) and no compensation (red-cross marker). For each pointing direction and each compensation configuration, four different polarization states are used, taking as reference the zero pointing. Error angles are plotted in logarithmic scale, $10\cdot\log_{10}\left(\Delta\epsilon\right)$, for clearer visualization of error angles close to $0$ rad. When the system performs the compensation, the error angle is below $0.2$ rad.}
\label{Fig:AngErrorRad}
\end{figure}

Figure \ref{Fig:AngErrorRad} shows the error angles measured with the receiving PL, separated in columns by pointing directions. They are identified in color markers for compensated and uncompensated system operation. Blue-circle marker for Tx-Rx and galvo compensation, green-diamond marker for galvo compensation, cyan-square marker for Tx-Rx compensation or red-cross marker for no compensation. Error angles are plotted in logarithmic scale, $10\cdot\log_{10}\left(\Delta\epsilon\right)$, for clearer visualization of error angles close to $0$ rad.

From Figure \ref{Fig:AngErrorRad} we see that the error angle due to the galvo for different pointing directions is small. In contrast, the error angle when not compensating for the Tx-Rx rotation angle is directly proportional to $\chi$. When the system performs the compensation, the error angle is smaller than $0.2$ rad. The main contribution to the error angle when compensating is due to system calibration loss due to drifts both in the SMF fibers and PC, and small pointing misalignment between the Tx and Rx along the line of propagation due to the manual placement of the PL. The performance appears to be limited by hysteresis in the PC actuators, which is largely but not completely cancelled by our min-to-max driving strategy. A driving strategy based on a tracking control of hysteretic piezoelectric actuator using adaptive rate-dependent controller \cite{Tan2009} could reduce this error significantly. We believe that $0.1$ is probably the lower limit given by the typical DOP degradation of commercial PCs. Typically, PC performance is limited by the polarization dependent loss (PDL) to about $0.1$ rad, while other parameters such as state-of-polarization (SOP) resolution and accuracy are well below this limit, $< 0.01$ rad and $< 0.002$ rad respectively. A common performance parameter in quantum communication applications is the quantum bit error ratio (QBER). QBER is defined as the number of correct sorted qubits to the number of detected qubits in the proper measuring bases, thus correct and wrong sorted qubits \cite{Gisin2002,Nielsen2010}. The relation between the QBER and the error angle follows a squared cosine function as QBER$=1-\cos^{2}\left(\Delta \epsilon\right)$. The upper error angle of $0.2$ rad corresponds to $3.95$\% QBER, while an improved driving strategy could achieve $0.1$ rad or $1$\% QBER. Most quantum protocols run with QBERs lower than $11$\%, thus the performance achieved is compliant with quantum key distribution protocols and validates the system to be used in free-space quantum communication links.

\section{Conclusions}
We have demonstrated a steerable optical system based on a feed-forward control with decoupled control of the polarization state. It is possible to assemble the system from commercially available components. The system is able to compensate the polarization controller particular hysteresis cycles and initial loading behavior, the particular galvo polarization transformation which depends on the pointing direction and the relative orientation rotation of the receiver.

The system demonstrated here enables to direct optical beams to a desired direction without affecting its polarization state or, if required, to perform a change of the polarization state as required by the receiver properties. The algorithm developed computes the voltages to apply to the polarization controller provided the two galvo mirrors angles and the relative Tx-Rx orientation rotation angle. With a simple hysteresis correction, the polarization error angle while steering the optical beam over the working field of view is smaller than $0.2$ rad. More accurate hysteresis compensation could reduce this error further. The experimental compensation was carried out in less than $\sim 20$ ms. This response time is probably limited by the mechanical response time of the galvanometer.

Although this demonstration used a transmitter and a receiver located in different places, we note that our polarization control solution is also applicable in receiver-less applications such as polarization optical coherence tomography, where transmitter and receiver share many elements of a common optical system. As there is no fundamental obstacle for the integration of the optical and mechanical components used here, robust integrated control systems can be built using this control strategy.

\section*{Acknowledgments}
This work was carried out with the financial support of Ministerio de Educaci\'on y Ciencia (Spain) through grants TEC2010-14832, FIS2007-60179, FIS2008-01051, FIS2010-14831 and FET-Open grant number: 255914 (PHORBITECH).

\bibliographystyle{IEEEtran}
\bibliography{PolConCompensationGalvoScanner}

\end{document}